\documentclass[aps,prl,twocolumn,amsmath,amssymb,nofootinbib,showpacs]{revtex4}
\newcommand{\figsize}{24.em}

\usepackage[english]{babel}
\usepackage{latexsym}
\usepackage{graphics}
\usepackage{epsfig}
\usepackage{color}
\usepackage{ulem}
\usepackage{mathrsfs}
\usepackage{hyperref}
\hypersetup{
colorlinks=true,
citecolor=blue,
linkcolor=red,
urlcolor=black
}

\newcommand{\epsfboxmod}[1]{\epsfbox{#1.eps}}
\newcommand{\infig}[2]{\begin{center}
                                    \mbox{ \epsfxsize #1 \epsfboxmod{#2}}
                                    \end{center}}

\newcommand{\ie}{{i.e.}}
\newcommand{\eg}{{e.g.}}
\newcommand{\etal}{{\it et al.}}

\newcommand{\be}{\begin{equation}}
\newcommand{\ee}{\end{equation}}

\newcommand{\vect}[1]{\boldsymbol{#1}}

\newcommand{\remove}[1]{}

\newcommand{\av}[1]{\overline{#1}}

\newcommand{\Vtrap}{V_\textrm{\tiny T}}
\newcommand{\Vr}{V_\textrm{\tiny R}}
\newcommand{\sigmar}{\sigma_\textrm{\tiny R}}

\newcommand{\kB}{k_{\textrm{\tiny B}}}

\begin{document}

\title{Localized and Extended States in a Disordered Trap}

\author{Luca Pezz\'e}
\author{Laurent Sanchez-Palencia}
\affiliation{
Laboratoire Charles Fabry de l'Institut d'Optique,
CNRS and Univ.~Paris-Sud,
Campus Polytechnique,
RD 128,
F-91127 Palaiseau cedex, France}

\date{\today}

\begin{abstract}
We study Anderson localization in a disordered potential combined with
an inhomogeneous trap.
We show that the spectrum displays both localized and extended states,
which coexist at intermediate energies.
In the region of coexistence,
we find that the extended states result from confinement by the trap
and are weakly affected by the disorder.
Conversely, the localized states correspond to eigenstates of the disordered potential,
which are only affected by the trap via an inhomogeneous energy shift.
These results are relevant to disordered quantum gases and
we propose a realistic scheme to observe the coexistence of localized and extended
states in these systems.
\end{abstract}

\pacs{03.75.-b, 03.75.Ss, 72.15.Rn}

\maketitle

Disorder underlies many fields in physics,
such as
electronics, superfluid helium and optics~\cite{Mott,reppy1992,akkermans2006}.
It poses challenging questions,
regarding quantum transport~\cite{phystoday2009} and
the interplay of disorder and interactions~\cite{disoint}.
In this respect, ultracold gases offer exceptionally well controlled
simulators for condensed-matter physics~\cite{reviewCMUA} and
are particularly promising for disordered systems~\cite{lsp2010}.
They recently allowed for the direct
observation of one-dimensional (1D) Anderson localization
of matter waves~\cite{damski2003,lsp2007,billy2008,roati2008}.
It should be noticed however that
ultracold gases do not only mimic standard models of condensed-matter physics,
but also raise new issues which require special analysis in its own right.
For instance, they are most often
confined in spatial traps, which has significant consequences.
On the one hand, retrieving information about
bulk properties requires specific algorithms~\cite{ho2010}.
On the other hand, trapping induces novel effects, such as
the existence of Bose-Einstein condensates in low dimensions~\cite{petrov2000},
and suppression of quantum tunneling
in periodic lattices~\cite{Pezze_2004}.

Consider Anderson localization~\cite{anderson1958}.
In \textit{homogeneous} disorder, linear waves can localize
owing to coherent multiple scattering,
with properties depending on the system dimension and the disorder strength~\cite{Mott}.
A paradigm of Anderson localization is that localized and extended states generally do not coexist in energy.
This relies on Mott's \textit{reductio ad absurdum}~\cite{Mott}:
Should there exist a localized state and an extended state with infinitely close energies
for a given configuration of disorder,
an infinitesimal change of the configuration would hybridize them,
forming two extended states.
Hence, for a given energy, almost all states should be either localized or extended.
Exceptions only appear for peculiar models of disorder
with strong local symmetries~\cite{DisoLocSymm}.
Then, a question arises:
Can \textit{inhomogeneous} trapping modify this picture 
so that localized and extended states coexist in energy?

In this Letter, we study localization in a disordered potential
combined with an inhomogeneous trap.
The central result of this work is the coexistence, 
at intermediate energies, of two classes of eigenstates.
The first class corresponds to states which spread over the full
(energy-dependent) classically allowed region of the bare trap,
and which we thus call ``extended''.
The second class corresponds to states of width much smaller
than the trap size, which are localized by the disorder,
and which we thus call ``localized''.
We give numerical evidence of
the coexistence of extended and localized states for different kinds of traps.
We show that
while the extended states are confined by the trap and weakly affected by the disorder,
the localized states correspond to eigenstates of the disordered potential,
which are only affected by the trap via an inhomogeneous energy shift.
Finally, we propose an experimentally-feasible
scheme using energy-selective time-of-flight (TOF)
techniques to observe this coexistence
with ultracold Fermi gases.

Let us consider a $d$-dimensional gas of noninteracting
particles of mass $m$,
confined into a spatial trap $\Vtrap(\vect{r})$
and subjected to a homogeneous disordered potential $V(\vect{r})$
of zero average,
amplitude $\Vr$ and correlation length $\sigmar$.
Hereafter, we use ``red-detuned'' speckle potentials ($\Vr<0$),
which are relevant to
quantum gases~\cite{lsp2010,noteREDvsBLUE}.
For the trap, we take $\Vtrap(\vect{r})=(\hbar^2/2ma^2) \vert\vect{r}/a\vert^\alpha$, 
being $a$ the trap length scale.
For instance, 
$\alpha=\infty$ and $a=L/2$
for a homogeneous box of length $L$,
while
$\alpha=2$ and
$a=\sqrt{\hbar/m\omega}$ for a harmonic trap of angular frequency $\omega$.
We numerically compute the eigenstates $\vert\psi_n\rangle$ and
eigenenergies $E_n$ of the Hamiltonian
\begin{equation}
\hat{H} = -{\hbar^2 \vect{\nabla}^2}/{2m} + V(\vect{r}) + \Vtrap(\vect{r}).
\label{eq:hamiltonian}
\end{equation}
The eigenstates are characterized by their center of mass,
$\vect{r}_n \! \equiv \! \langle \psi_n \vert \hat{\vect{r}} \vert \psi_n\rangle$,
and
spatial extension (rms size),
$\Delta {r}_n \! \equiv \! \left( \langle \psi_n \vert \hat{\vect{r}}^2 \vert \psi_n\rangle
                            \! - \! \vect{r}_n^2 \right)^{1/2}$.
The quantity $\Delta {r}_n$ quantifies localization:
the smaller, the more localized.


\begin{figure*}[!t] 
\begin{center} 
\includegraphics[scale=0.22]{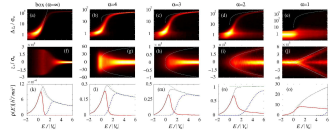}
\end{center}
\vspace{-0.5cm}
\caption{Extension {(a)-(e)}, 
center of mass (f)-(j) and 
density of states [DOS] (k)-(o)
of the eigenstates versus energy in various kinds of 1D disordered traps.
The plots result from accumulation of numerical data over 5000
realizations of a red-detuned speckle potential with
$m\sigmar^2\vert\Vr\vert/\hbar^2 \! = \! 0.256$.
The first column refers to a flat box of length $L \! = \! 500\sigmar$.
The curved line in (a) corresponds to an infinite system~\cite{noteFINITE}
and the horizontal line is $\Delta z^0 \! = \! L/2\sqrt{3}$.
The other columns refer to inhomogeneous traps
with $a=12.5\sigmar$ and various trap powers $\alpha$.
The solid lines correspond to the nondisordered case,
\ie\ $\Delta z^0$ in panels~(b)-(e) and $\pm z_\textrm{cl}(E)$ in panels~(g)-(j).
The last row shows the full DOS $\rho(E)$ (solid black line),
as well as the DOS restricted to
localized ($\rho_<$, solid red line)
and
extended ($\rho_>$, dashed blue line)
states~\cite{noteSEGREG}.
The dot-dashed green lines are the nondisordered limits.
}
\label{fig:coexistence}
\end{figure*}

Numerical results for the
1D (d=1) case are reported in Fig.~\ref{fig:coexistence}.
In infinite, homogeneous disorder {($\alpha=\infty$, $L=\infty$)},
all states $\vert \psi_n\rangle$ are localized,
uniformly distributed in space,
and, for most models of disorder, their extension  $\Delta z_n$ monotonically 
increases with the energy~\cite{lifshits1988}.
As Figs.~\ref{fig:coexistence}(a),(f) show, 
using a finite flat box ($L<\infty$)
only induces a trivial finite-size effect:
For low-enough energy $E$, we find
$\Delta z_n \ll L$
and the states are not significantly affected by the finite size of the box.
For larger energies however, boundary effects come into the picture.
The states are centered close to the box center
and their extension saturates to the value
obtained for a plane wave, \ie\ $\Delta z^0 \! = \! L/2\sqrt{3}$.
A central outcome of these results is that the curve giving
$\Delta z_n$  versus $E$ displays a single branch.
In particular, there is no energy window where localized and extended states coexist.
This finding holds independently of the finite box size
and is in agreement with Mott's argument~\cite{Mott}.

For inhomogeneous traps ($\alpha<\infty$),
we find a completely different behavior.
The curves giving $\Delta z_n$ and $z_n$ versus $E$
now display two clearly separated
branches
[see Figs.~\ref{fig:coexistence}(b)-(e) and (g)-(j)].
For low energy, the states are strongly localized and,
for $E>0$, they are roughly uniformly distributed in a region
bounded by the (energy-dependent) classical turning points,
$z_\textrm{cl}(E)$, defined as the solutions of
$\Vtrap(z_\textrm{cl})=E$.
For higher energy,
the extension of the states corresponding to the upper branch
in Figs.~\ref{fig:coexistence}(b)-(e) grows and eventually saturates to that of
the eigenstates of the nondisordered trap,
$\Delta z^0 (E)$.
The centers of mass of these states 
approach the trap center and form the horizontal branch
in Figs.~\ref{fig:coexistence}(g)-(j).
This branch corresponds
to extended states. It is easily interpreted 
in terms of finite-size effects, similarly as for a finite flat box.
The lower branch in Figs.~\ref{fig:coexistence}(b)-(e) is more surprising.
It identifies strongly localized states of relatively large energy.
It has no equivalent in the flat box
and cannot be interpreted as a finite-size effect.
The corresponding states are located close to
the classical turning points $z_\textrm{cl}(E)$
and generate the outer branches in
Figs.~\ref{fig:coexistence}(g)-(j).
As Fig.~\ref{fig:coexistence} shows, this holds for all inhomogeneous traps.
When the trap power $\alpha$ increases,
the branch of extended states gets denser at the expense of that of localized states,
and completely vanishes for $\alpha=\infty$ (homogeneous box).

\begin{figure}[!t] 
\begin{center} 
\infig{\figsize}{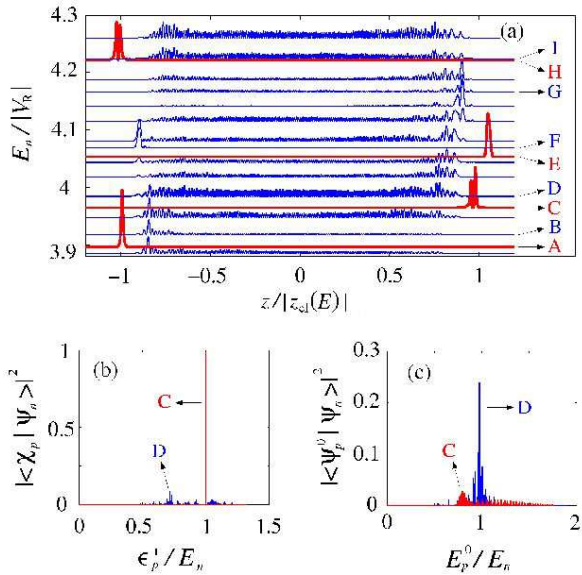}
\end{center} 
\vspace{-0.4cm}
\caption{\small{
Eigenstates for a single realization of a 1D disordered harmonic trap.
(a) Non-normalized spatial densities, $\vert\psi_n(z)\vert^2$,
vertically displaced to their eigenenergy $E_n$.
Thick red lines correspond to localized states,
and thin blue lines to extended states~\cite{noteSEGREG}.
Note that extended and localized states may occupy almost-degenerate energy levels
(\eg\ H and I).
The states C and D are projected:
(b) over the eigenstates of the disordered potential, $\vert \chi_p \rangle$, and 
(c) over those of the harmonic trap, $\vert \psi_p^0 \rangle$.
The parameters are as in Fig.~\ref{fig:coexistence}.}}
\label{fig:project}
\end{figure}

The coexistence of
localized and extended states in the same energy window
for disordered traps
is confirmed on more quantitative grounds
in the last row of Fig.~\ref{fig:coexistence}. It shows
the full density of states (solid black line),
as well as the density of
localized ($\rho_<$, solid red line)
and extended ($\rho_>$, dashed blue line)
states~\cite{noteSEGREG}.
The different nature of the localized and extended states is
even more striking when one studies the wavefunctions.
Let us focus for instance on the harmonic trap ($\alpha=2$) and on a
narrow slice of the spectrum around $E \sim 4 \vert\Vr\vert$,
where $\rho_</\rho \simeq 14\%$ of the states are localized~\cite{noteOtherTraps}.
Figure~\ref{fig:project}(a) shows the spatial density $\vert\psi_n (z)\vert^2$
of all states found for a \textit{single realization} of the disorder.
We can clearly distinguish localized (thick red lines) and extended (thin blue lines)
states, which shows that they coexist in the same energy window
for each realization of the disorder.
The localized states are very narrow and present
no node (\eg\ states A and E) or a few nodes (\eg\ states C and H).
They may be identified as bound states of the local deep wells of the disordered potential,
similarly as the lowest-energy states creating the Lifshits tail
in bare disorder~\cite{lifshits1988}.
To confirm this, let us decompose the eigenstates 
$\vert\psi_n\rangle$ of the disordered trap
onto the basis of the eigenstates $\vert\chi_p\rangle$
of the bare disordered potential [\ie\ Hamiltonian~(\ref{eq:hamiltonian}) with $\Vtrap \equiv 0$],
associated to the eigenenergies $\epsilon_p$.
For a localized state $\vert\psi_n\rangle$,
we find $\vert \langle \chi_p \vert \psi_n \rangle \vert^2\sim 1$
for a single state $\vert\chi_p\rangle$
such that $\epsilon^{\prime}_p \simeq E_n$,
where $\epsilon^{\prime}_p = \epsilon_p + \langle\chi_p\vert \Vtrap (z)\vert\chi_p\rangle$
is the eigenenergy of $\vert\chi_p\rangle$ shifted by the trapping potential
[see Fig.~\ref{fig:project}(b)].
Conversely, the same decomposition for an extended state
shows a broad distribution of amplitude much smaller than unity.
A localized state $\vert \psi_n\rangle$ of the disordered trap thus corresponds to
a strongly localized state $\vert \chi_p\rangle$ in the bare disorder,
which is affected by the trap by just the energy shift
$\langle\chi_p\vert \Vtrap (z)\vert\chi_p\rangle$.
We generally find that
$\vert\epsilon_p\vert \ll \langle\chi_p\vert \Vtrap (z)\vert\chi_p\rangle
\simeq \langle\psi_n\vert \Vtrap (z)\vert\psi_n\rangle$,
and, due to the reduced spatial extension of $\vert\psi_n\rangle$,
we get $E_n \simeq \Vtrap(z_n)$~\cite{note:stateC}.
This explains that the localized states are located close to the classical turning points,
as observed in Figs.~\ref{fig:coexistence} and \ref{fig:project}(a).

Let us now decompose the states $\vert\psi_n\rangle$ of the disordered trap
onto the basis of the eigenstates $\vert\psi^0_p\rangle$ of the bare trap
[\ie\ Hamiltonian~(\ref{eq:hamiltonian}) with $V = 0$], associated to the eigenenergy $E^0_p$.
For a localized state, the distribution is broad.
Conversely, for an extended state, the distribution is sharp
and peaks at $E^0_p \simeq E_n$ to a value equal to a fraction of unity
[see Fig.~\ref{fig:project}(c)].
An extended state may thus be seen as reminiscent of an eigenstate of the bare trap,
which is weakly affected by the disorder.
Still, the main peak in Fig.~\ref{fig:project}(c) is smaller than unity.
Only for significantly higher energy,
the state $\vert\psi_n\rangle$ results from weak perturbation
of $\vert\psi_n^0\rangle$,
and $\vert \langle \psi_p^0 \vert \psi_n \rangle \vert^2$
displays a main peak of the order of unity
as predicted by standard perturbation theory.

Our results can now be easily interpreted.
In bare disorder, the typical size $\Delta z$ of the localized states increases
faster than the classically allowed region $z_\textrm{cl} \propto E^{1/\alpha}$
provided by the trap.
For low energy, $\Delta z \ll z_\textrm{cl}$
so that the states are strongly localized by the disorder and weakly affected by the trap.
For higher energy however, the disorder would localize the states on a scale
exceeding $z_\textrm{cl}$.
The states are then bounded by the trap
and the effect of disorder becomes small.
This forms the branch of extended states in both the disordered box and traps.
In addition, some strongly localized states
with very low energy in the bare disorder and
located around point $z_n$ are shifted by the trap
to approximately the energy $\Vtrap (z_n)$.
This forms the branch of localized states
only in disordered traps ($\alpha < \infty$)
since, in the box, a state cannot be placed at intermediate energy
due to the infinitely sharp edges.
Quantitatively, since the localized states in the bare disorder are uniformly distributed
in space, the density of localized states can be estimated to roughly scale as
$\rho_< \propto ({1}/{\alpha})\times E^{1/\alpha-1}$,
which is consistent with the disappearance of the branch of localized states
when $\alpha$ grows and with its vanishing for $\alpha = \infty$ (see Fig.~\ref{fig:coexistence}).
Still, it is striking that localized and extended states can coexist
in the same energy window.
The disordered potential combined with a \textit{smooth} trap
permits localized states to sit outside the classically allowed region
occupied by extended states [see Fig.~\ref{fig:project}(a)].
Then, the Mott argument does not apply here
because the spatial segregation can be strong enough to 
suppress hybridization
for an infinitesimal change of the disorder configuration.

\begin{figure}[!t] 
\begin{center} 
\infig{\figsize}{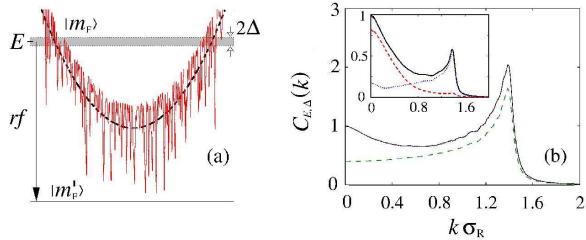}
\end{center} 
\vspace{-0.4cm}
\caption{
Scheme to observe the coexistence of localized and extended states
in disordered traps (solid red line).
(a) Atoms occupying the eigenstates
of energy $E \pm \Delta$ (shaded region) are transfered
to a different internal state via rf coupling.
The corresponding momentum distribution is then measured by TOF.
(b) Correlation function $\mathcal{C}_{E,\Delta}(k)$ (black solid line)
and momentum distribution $\av{\mathcal{D}_{E,\Delta}(k)}$
(dashed green line, arbitrary units), for $\Delta = 2 \hbar\omega$.
Inset: $\mathcal{C}_{E,\Delta}(k)$ of all states (solid black line),
and separating localized (dashed red line)
and extended (dotted blue line) states~\cite{noteSEGREG},
for $\Delta = 0.01\hbar\omega$.
Here $E = 4\vert\Vr\vert$ and the other parameters are as in Fig.~\ref{fig:coexistence}.}
\label{fig:correlation}
\end{figure}

Let us now discuss a possible scheme to observe
the coexistence of localized and extended states in a disordered trap.
Consider a gas of noninteracting ultracold fermions
prepared in a given internal state, at temperature $T$ and chemical potential $\mu$.
A class of energies $\vert E_n-E \vert \lesssim \Delta$
[see Fig.~\ref{fig:correlation}(a)]
deep in the Fermi sea (\ie\ with $\mu-E \ll \kB T$)
can be selected by applying a spin-changing radio-frequency (rf)
field of frequency $\nu=E/h$ and duration $\tau \sim h/\Delta$
(with $h$ the Planck constant)~\cite{greiner2003,Pezze_2004,Guerin_2006}.
The rf field transfers the atoms of corresponding energies 
to an internal state insensitive to the disordered trap.
The transfered atoms expand freely,
which provides their momentum distribution:
\begin{equation}
\mathcal{D}_{E,\Delta}(k) ~~~ \simeq \sum_{\vert E_n \! - \! E \vert \lesssim \Delta}
\vert \hat{\psi}_n (k) \vert^2,
\label{eq:momentum}
\vspace{-0.3cm}
\end{equation}
where $\hat{\psi}_n (k)$ is the Fourier transform of $\psi_n (z)$
[TOF technique].
In the coexistence region, $\mathcal{D}_{E,\Delta}(k)$ has two
significantly different contributions:
For localized states, $\vert\hat{\psi}_n (k)\vert^2$ is centered around $k \simeq 0$
with tails of width $\Delta k_n \sim \Delta z_n^{-1}$.
Conversely, for extended states, $\vert\hat{\psi}_n (k)\vert^2$ is peaked at $k \simeq \sqrt{2mE}/\hbar$
with long tails towards small momenta.
We however found that averaging over realizations of the disorder
blurs the central peak associated to the localized states
in $\av{\mathcal{D}_{E,\Delta}(k)}$.
In turn, the quantity
$\mathcal{C}_{E,\Delta}(k) \equiv 
\av{\mathcal{D}_{E,\Delta}(k) \times \mathcal{D}_{E,\Delta}(0)} / 
\av{\mathcal{D}_{E,\Delta}(0)^2}$
displays two distinct peaks
for a rf pulse of realistic durations [see Fig.~\ref{fig:correlation}(b)].
The central one is more pronounced for narrower pulses.
Selecting either the localized states or
the extended states~\cite{noteSEGREG}
confirms that the central peak corresponds to the localized states
and the side peak to the extended states
[see Inset of Fig.~\ref{fig:correlation}(b)].

\begin{figure}[!t] 
\begin{center} 
\infig{\figsize}{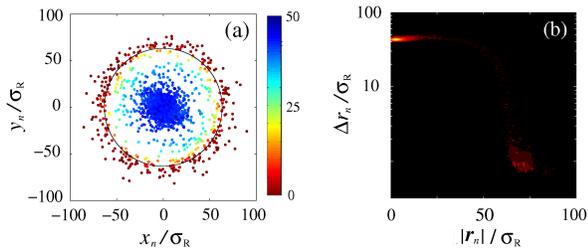}
\end{center} 
\vspace{-0.4cm}
\caption{ 
Coexistence of localized and extended states in a 2D disordered harmonic trap
for $E/\vert\Vr\vert = 4\pm 0.0003$.
The figure results from accumulation of data from $2\times 10^4$ realizations
of the disorder,
with $m\sigmar^2\vert\Vr\vert/\hbar^2=0.8$ and  
$\omega=0.05\vert\Vr\vert/\hbar$.
(a) Centers of mass $\vect{r}_n$ of the eigenstates
and corresponding values of $\Delta {r}_n/\sigmar$ in color scale.
The solid black line is the line of classical turning points,
$r_\textrm{cl}(E)=\sqrt{2E/m\omega^2}\simeq 63.2\sigmar$.
(b) Extension $\Delta {r}_n$ versus distance from the trap center $\vert\vect{r}_n\vert$.}
\label{fig:twoD}
\end{figure}

Finally,
we have performed similar calculations as above in a 2D harmonic trap.
Figure~\ref{fig:twoD}(a) shows the centers of mass $\vect{r}_n$
of the eigenstates with $E_n \simeq 4\vert\Vr\vert$,
the color scale giving $\Delta {r}_n$.
Figure~\ref{fig:twoD}(b) shows a density plot of $\Delta {r}_n$ versus $\vert\vect{r}_n\vert$
for the same data.
Again, the eigenstates clearly separate into two classes:
Some states are extended (large $\Delta {r}_n$) and centered nearby the trap center
(small $\vert\vect{r}_n\vert$).
The other states are strongly localized (small $\Delta {r}_n$)
and located nearby the line of classical turning points
($\vert\vect{r}_n\vert \simeq r_\textrm{cl}(E)=\sqrt{2E/m\omega^2}$).
Hence, the two classes of states can coexist at intermediate energies also
in 2D disordered traps.

In conclusion, we have shown that,
in a disordered inhomogeneous trap,
localized and extended states can coexist
in a given energy window.
The localized states correspond to eigenstates of the disordered potential
which are only affected by the trap via an inhomogeneous energy shift.
Conversely, the extended states spread over 
the classically allowed region of the trap and are weakly affected by the disorder.
This effect is directly relevant to presentday experiments with disordered quantum gases,
which are most often created in harmonic traps 
\cite{roati2008,clement2008,chen2008,pezze2009,deissler2010}.
We have proposed a realistic scheme to observe it in these systems.
In the future, it would be interesting to extend
our results to higher dimensions and to other kinds of inhomogeneous disordered systems.

\acknowledgements
We thank T. Giamarchi and B. van Tiggelen for discussions and
ANR (Contract No.\ ANR-08-blan-0016-01),
Triangle de la Physique, LUMAT and IFRAF
for support.


\begin{thebibliography}{99}
\bibitem{noteREDvsBLUE}
We obtained the same qualitative behavior as reported in the Letter
using other models of disorder,
such as ``blue-detuned'' speckle potentials and
Gaussian impurities.
\bibitem{noteSEGREG}
We use the {condition $\Delta z \! < \! 10\sigmar$}
{to identify localized states,
and $\Delta z \! > \! (3/4)\Delta z^0(E)$ for extended states}.
\bibitem{noteOtherTraps}
{We obtained qualitatively-similar results for different energies and different traps.}
\bibitem{note:stateC}
For instance, we find
$E_n \simeq 3.97 \vert \Vr \vert$,
$\epsilon_p \simeq 0.24 \vert \Vr \vert$ and
$\langle \chi_p \vert \Vtrap \vert \chi_p \rangle \simeq 3.71 \vert \Vr \vert$
for state C in Fig.~\ref{fig:project}.
\bibitem{noteFINITE}
Here, we use a box
of length $3000\sigmar$
much larger than $\Delta z$ for all states considered.
\end{thebibliography}
\end{document}